# Disappearing of the Fermi level pinning at semiconductor interfaces

Jin-Peng Yang[1*], Nobuo Ueno[2*]

*1, College of Physical Science and Technology, Yangzhou University, 225009, China;*

*2, Graduate School of Advanced Integration Science, Chiba University, Chiba, 2638522, Japan*

\*Correspondence to: yangjp@yzu.edu.cn, uenon@faculty.chiba-u.jp

**Abstract**

We identify a universality in the Fermi level change of Van der Waals interacting semiconductor interfaces-based Schottky junctions. We show that the disappearing of quasi-Fermi level pinning at a certain thickness of semiconductor films for both intrinsic (undoped) and extrinsic (doped) semiconductors, over a wide range of bulk systems including inorganic, organic, and even organic-inorganic hybridized semiconductors. The Fermi level ($E_F$) position located in the energy bandgap was dominated by not only the substrate work function ($\Phi_{sub}$), but also the thickness of semiconductor films, in which the final $E_F$ shall be located at the position reflecting the thermal equilibrium of semiconductors themselves. Such universalities originate from the charge transfer between the substrate and semiconductor films after solving one-dimensional Poisson's equation. Our calculation resolves some of the conflicting results from experimental results determined by using ultraviolet photoelectron spectroscopy (UPS) and unifies the general rule on extracting $E_F$ positions in energy bandgaps from (i) inorganic semiconductors to organic semiconductors and (ii) intrinsic (undoped) to extrinsic (doped) semiconductors. Our findings shall provide a simple analytical scaling for obtaining the 'quantitative energy diagram' regarding thickness in the real devices, thus paving the way for a fundamental understanding of interface physics and designing functional devices.

Contacting a substrate with semiconductor materials to form a substrate-semiconductor contact is an inevitable process for any semiconductor device. [1-4] The contact often leads to the formation of an injection barrier (or Schottky barrier) and band bending, in which both of these two factors dominate the electron/hole injection efficiency and device performance.[5-8] The Schottky barrier height ($\Phi_B$), which is defined by the energy difference between substrate work function ($\Phi_{sub}$) and the electron affinity (or ionization potential) of the semiconductor, is followed by (i) a Schottky-Mott rule (using interface S parameter with S=1) and (ii) a quasi-Fermi level pinning (S=0). [9-11] On the other hand, the built-in potential energy ($\Phi_{bi}$) in the band bending region is extracted by the energy difference between $\Phi_{sub}$ and semiconductor work function ($\Phi_{semi}$), which provides the information of the width of space charge and influences carriers' injection/separation efficiency. [12] These interfacial regions governed by electrostatics result in a change of $\Phi_B$ and $\Phi_{bi}$ that have been actively studied due to their broad applications existing in inorganic semiconductor devices and expanding to organic and organic-inorganic hybridized semiconductors within recent years. [13, 14]

Experimentally, enormous effects have been performed to reveal the energy diagram either with direct detections (i.e., UPS, inverse photoemission spectroscopy, and Kelvin probe methods) [15-17] or indirect methods (i.e., extracting data from relevant devices) [18, 19]. However, experimental results give diverse conclusions on the Fermi level ($E_F$) positions regarding semiconductor materials. Taking $E_F$ in inorganic semiconductor as an example, the Schottky-Mott model (S=1) is rarely obtained due to the existence of several types of chemical interactions that are difficult to avoid at the interface, [20, 21] although a nonrectifying Ohmic contact could be realized (if $\Phi_B \ll k_BT$). In contrast to inorganic semiconductors, the junction formed at organic semiconductors following a Schottky-Mott model (S=1) could be easily found, while rarely forming a nonrectifying Ohmic contact ($\Phi_B$ is commonly in-between of 0.2-0.6 eV). [22-24] Another aspect, $\Phi_{bi}$, depicts a total energy change in the band bending region, is another critical factor and should be carefully considered. $\Phi_{bi}$ directly reflects the magnitude of internal electric field and the width of depletion layer at semiconductor interfaces. The latter is paramount important for precisely probing $E_F$ positive in energy bandgaps since it provides (i) a critical thickness to reach the thermal equilibrium and (ii) the necessary thickness for excluding band structure evolution at interfaces to study the doping efficiency of extrinsic semiconductors. Experimentally, however, the width of band bending is commonly considered to be extended from tens of nanometers to over several hundred nanometers depending on (i) different materials (organic or inorganic) and (ii) various doping ratios (doped or undoped) and lacking the university. [4, 15, 25, 26] Although various models have been proposed to interpret the experimental results and to elucidate the key factors governing $\Phi_B$ and $\Phi_{bi}$ according to different types of semiconductors, a consistency remains unclear.

In this Letter, we develop calculations for precisely giving $E_F$ position to quantitatively elucidate the generalized model of $\Phi_B$ and $\Phi_{bi}$ over different types of Van der Waals interacting semiconductor interfaces, including organic and inorganic, intrinsic and extrinsic systems. The key findings are threefold. First, we elucidate a universal phenomenon on probing $E_F$ position at various semiconductor interfaces, which shows that the disappearance of quasi-Fermi level pinning at a certain thickness of semiconductor films for both intrinsic (undoped) and extrinsic (doped) semiconductors. The $E_F$ position located in the energy bandgap was governed by (i) the $\Phi_{sub}$, and (ii) the thickness of semiconductor films, in which the latter is often ignored in semiconductor interfaces. Second, the total amount of transferred charges at exact contact interfaces does not simply follow the density of dopants but is strongly related to $\Phi_{sub}$. The total transferred charges could show an extremely high density than the dopants, which indicates the generalized model only considering density of dopants in these interfaces is not valid. Third, we have quantitatively revealed that the band bending region could reach as high as millimeters to centimeters for intrinsic semiconductors. However, it rapidly shrinks to tens of nanometers after doping, giving that the thickness evolution of band bending changes as an exponential type. This finding further suggests that commonly observed band structures for different semiconductor interfaces via surface probe methods could not exclude charge transfer effects from contacted substrates. Thus, our results resolve some of the conflicting results from experimental results resolved by using UPS and unifying the general rule related to probe $E_F$ position in energy bandgaps either for intrinsic or extrinsic semiconductors.

We consider two types of density of states (DOS) distributions [D(E)] with one is Parabolic-dispersion band [$D(E) \propto E^{1/2}$] for a model of three dimensional crystal semiconductors and the other one is a Gaussian DOS for disordered semiconductors [e.g., amorphous film with $D(E) \propto exp(-E^2)$]. The illustrations on two types of DOS distributions with detailed DOS distribution functions of $D_C(E)$, $D_V(E)$, $D_H(E)$, and $D_L(E)$ are independently given in supplementary materials (**SM**). For an ideal intrinsic (undoped) semiconductor, the concentration of electrons (n) and holes (p) are given as:

$$n = \int D_{C(L)}(E) \cdot f(E) \cdot dE \qquad (1)$$

$$p = \int D_{V(H)}(E) \cdot [1 - f(E)] \cdot dE \qquad (2)$$

Where $f(E)$ is the Fermi-Dirac distribution function. After considering the thermal equilibrium condition (p = n) and using parameters listed in **Table I**, it is found that the intrinsic $E_F$ position should be located in the middle of energy bandgap [27] for both types of DOS (seen in **Figure S1**).

For an extrinsic (doped) semiconductor, the dopants either with acceptors ($N_A$) or donors ($N_D$) would shift $E_F$ close to the valence band ($E_{V/H}$) or the conduction band ($E_{C/L}$) by obeying the charge neutrality (p + $N_D^+$ = n + $N_A^-$) with the following Equations:

$$p + N_D^+ = \int D_{V(H)}(E) \cdot [1 - f(E)] \cdot dE + N_D / \{1 + g_D \cdot \exp[(E_F - E_D)/k_B T]\} \qquad (3)$$

$$n + N_A^- = \int D_{C(L)}(E) \cdot f(E) \cdot dE + N_A / \{1 + g_A \cdot \exp[(E_A - E_F)/k_B T]\} \qquad (4)$$

Where $E_D$, $E_A$, $g_D$ and $g_A$ are energies and degeneracy factors of donor and acceptor, $k_B$ is a Boltzmann constant, and T is the temperature (T=298 K).

Consider a substrate-semiconductor interface with a general condition under thermal equilibrium, solving one-dimensional Poisson's equation, with conditions (i) the consistency of vacuum level and (ii) the thermal equilibrium reached at a final position x, could reveal the change of electrostatic potential [V(x)] and electron potential energy [Φ(x)],

$$\nabla \cdot \nabla \Phi(x) = \nabla \cdot \nabla [-eV(x)] = -e\rho(x)/(\varepsilon_r \varepsilon_0) \quad (5)$$

Where $\rho(x) = e \cdot [p(x) + N_D^+(x) - n(x) - N_A^-(x)]$, represents a net charge density distribution, p(x) is $\int D_{V(H)}[E + \Phi(x)] \cdot [1 - f(E)]dE$, n(x) is $\int D_{C(L)}[E + \Phi(x)] \cdot f(E)dE$, e is the positive elementary charge, $\varepsilon_0$ and $\varepsilon_r$ are electric constant and permittivity. The calculated parameters used to demonstrate our exemplary results are summarized in **Table I**.

In the first step, we will show the influence of doping ratio ($N_A/N_{V(H)}$) and distance from the interface (x) on $E_F$ shift in the energy bandgap of semiconductors. **Figure 1 (a)** depicts the calculated results on $E_F$ position change with respect to the valence band edge ($E_V$ or $E_H$) under different doping ratios after solving **Eqs.** (1) ~ (4). The inset figure gives energy positions of the conduction band-CB (2.5 eV) and the valence band-VB (4.9 eV). We can see that $E_F$ of intrinsic semiconductors (denotes as $E_{Fi}$) is in the position of middle energy bandgap either for a square-root type DOS or a Gaussian type DOS, which is also consistent with our common understanding and agrees well with previous results. [27] On the other hand, $E_F$ depicts a distinguishable energy shift when the doping ratio of acceptors ($N_A/N_{V(H)}$) changes from $10^{-10}$ to $10^{-1}$ under the assumption of $E_A$ with 0.4 eV. For a square-root type DOS, $E_F$ shifts linearly in a semi-logarithm scale with a slope of $k_BT$ (0.026 eV at 298 K). In contrast to a square-root type DOS, $E_F$ demonstrates a nonlinear energy shift for a Gaussian type DOS, where a "tangent change" appears under a certain doping ratio. **Figures 1 (b)** and **(c)** give hole injection barrier ($\Delta_h = E_F - E_{V/H}$) change as a function of distance under a series of different doping ratios [extracted from **Figure 1 (a)**], where we treat DOS distributions with square-root type in (b) and Gaussian type in (c), and choose the $\Phi_{sub}$ with 3 eV (in the energy bandgap region). We can see the unchanged $E_F$ appears at the interface close to the substrate region both for the intrinsic and doped semiconductors (i.e., $N_A/N_V=10^{-8} \sim 10^{-3}$ for a square-root DOS, $N_A/N_H=10^{-5} \sim 10^{-2}$ for a Gaussian DOS). Whereas, unchanged $\Delta_h$ appears only as a temporary status versus distance, which is no longer available once the distance from the interface has been increased for all types of DOS.

For an intrinsic type shown in **Figure 1 (b)**, the $\varDelta_h$ initially stays in an energy value of 1.9 eV and keeps unchanged till the distance reaches ~$10^4$ nm. Then the $\varDelta_h$ shifts linearly in a semi-logarithm scale and finally reaches a stable energy value of 1.2 eV, which represents the middle energy bandgap seen in **Figure 1 (a)**. In contrast to the intrinsic case, unchanged $\varDelta_h$ can be only found at an interface within narrow regions (several nanometers) when semiconductors are heavily doped (i.e., $N_A/N_V=10^{-3}$), and then $\varDelta_h$ falls rapidly to reach a stable energy value (0.21 eV) within tens of nanometers. This unchanged $\varDelta_h$, also represents the energy position of 4.69 eV (4.9 eV -0.21 eV) shown in **Figure 1 (a)**, which implies reaching the neutral region and reflecting characteristics of semiconductors aside from the interface region. **Figure 1 (c)** gives the $\varDelta_h$ shift similar to **Figure 1 (b)**, although the different DOS distribution has been included.

The origin of an initially stable $\varDelta_h$ at the interface regions as shown in **Figure 1 (b)** and **(c)** would be elucidated in **Figure 2**, where it depicts the $\varDelta_h$ change as a distance function considering different $\Phi_{sub}$, representing intrinsic and doped semiconductors with DOS distribution as square-root types in (a) and Gaussian types in (b). The dopant ratios selected for calculations and comparisons are also listed in the Figure. For a square-root DOS shown in **Figure 2 (a)**, the following points can be found: (i) $\varDelta_h$ is dominated both by distance from the interface and $\Phi_{sub}$, where the change of $\varDelta_h$ is restricted to two red dash lines included by a triangle shape (light yellow shadow) and irrelevant to types of semiconductors (intrinsic or doped). (ii) Unchanged $\varDelta_h$ (with $E_F$ pinning) appears at boundaries (two red dash lines) and demonstrates independence of $\Phi_{sub}$. If $\Phi_{sub}$ is larger than $E_V$ (or smaller than $E_C$), the quasi-$E_F$ pinning (S=0) appears not only at interface regions (few nanometers) but also in the thicker region of semiconductors. However, if $\Phi_{sub}$ is in-between $E_V$ and $E_C$, $E_F$ position can be calculated from the energy difference between $E_V$ and $\Phi_{sub}$ (S=1) and keeps unchanged before reaching boundaries of those two red dash lines. (iii) Doping significantly shortens the distance to reach a neutral region (total interface region is narrowed). However, this change is still restricted inside the boundaries for intrinsic semiconductors (red dash lines). More importantly, $E_F$ pinning with S=0 could disappear at interface regions relying both on the distance from direct contacts and doping ratios. On the other hand, calculated results considering Gaussian DOS in **Figure 2 (b)** depicts a similar observation as shown in **Figure 2 (a)**, one main difference is still observed: the shift of $\varDelta_h$ for intrinsic semiconductors, in which the trend at boundaries changes from a linear relation in a semi-logarithm scale to a curved relation (twisted triangle shadow region). This is mainly due to the DOS difference between a square-root type and a Gaussian type since there is still some "extra-tailing" DOS for a Gaussian type DOS in the energy bandgap region, while no additional DOS exists for a square-root type.

Finally, comparisons of transferred charges with different doping ratios under various $\Phi_{sub}$ are given, which could allow us to obtain a quantitative understanding of origins of the $\varDelta_h$ shift. **Figure 3** shows exemplary charge density distributions (ρ) for calculated results of **Figure 2** with various doping ratios and $\Phi_{sub}$, where square-root type DOSs are given in **(a)** and **(c)** and Gaussian type DOSs are given in **(b)** and **(d)**. The calculated results for intrinsic types are given in **SM** (seen in **Figure S2**). Distributions of transferred charges along with the distance from contact (x= 0 nm) also demonstrate three similarities between square-root types and Gaussian types for doped semiconductors, as shown in **Figure 3**. First, a more considerable amount of transferred charges appears when $\Phi_{sub}$ is either higher than $E_{V(H)}$ or lower than $E_{C(L)}$. Notably, these charges assembled with extremely high amounts at directly contacted regions (a few nanometers to tens of nanometers) when $\Phi_{sub}$ is very low (i.e., 2 eV and 2.5 eV). This is significantly different considering what we have learned from semiconductor textbooks [4], where charge density is always assumed to be a constant in the whole metal-semiconductor junction region. Second, we could also see the constant charge density within a certain distance range, which indicates the induced band bending at these regions are due to the contributions from dopants. However, a maximum distance for charges could always be found if a doping ratio is settled, which is irrelevant to $\Phi_{sub}$ [see calculated results with $\Phi_{sub}$ of 2 and 2.5 eV in **Figure 3 (a)** and **(b)**]. Third, despite a square-root type or a Gaussian type DOS at semiconductor interfaces, the transferred charge does not decay rapidly in the region before reaching the final neutral position (ρ=0). As shown in **Figure 3 (a)** and **(b)**, a gradual change of charge density could extend with the length of hundreds of nanometers before reaching the "ideal" neutral region.

We further show two typical semiconductor interfaces to specify our calculations, in which one is perovskite material ($CH_3NH_3PbI_3$), and the other is classical inorganic silicon. **Figure 4 (a)** depicts the $\varDelta_h$ change as a function of film thickness in $CH_3NH_3PbI_3$, where different defects are considered based on experimental observations. [28, 29] The parameters obtained from experiments are also given in **Table I**. The ultralow density of gap states/defects, which has been experimentally reported (from single-crystal with $3\times10^{10}$ cm$^{-3}$ to polycrystalline with $3\times10^{14}$ cm$^{-3}$) [30, 31], could easily shift the $E_F$ to $E_C$ (larger $\varDelta_h$) and result in the probed spectra (i.e., UPS) with demonstrating N-type or even heavily doped N-type properties. Moreover, our calculation further implies the necessary thickness for $CH_3NH_3PbI_3$ to overcome the interfacial charge transfer region after contact. The thickness could extend from hundreds of nanometers to tens of micrometers depending on total amount of defects inside semiconductors. **Figure 4 (b)** shows calculated results on the mapping of $\varDelta_h$ for an ideal substrate-silicon junction. Since $\Phi_{sub}$ is chosen between silicon energy bandgaps, the unchanged $\varDelta_h$ with 0.48 eV could be expected at the interface region. On the other hand, the $E_F$ position will eventually return to the extrinsic position reflecting the dopants in silicon with increased thickness. A gradual decrease of $\varDelta_h$ could be observed due to increased dopants in the neutral region. Interestingly, a linear shift in the transition region (from an interface to a neutral) could be found and marked with a dashed line. This directly indicates that increased dopants in silicon could linearly reduce the thickness of depletion layers (also denoted as the width of charge transfer region), which is entirely different from a well-known relation from semiconductor textbooks ( $W \propto N_a^{-1/2}$ ). [4] A similar tendency could also be found at an organic semiconductor interface (pentacene), seen in **Figure S3**. Therefore, assuming constant charge density (dopants) to solve the Poisson equation and to further elucidate the energy diagram in substrate-semiconductor interfaces needs to be reconsidered.

In summary, we have demonstrated a universality in the Fermi level pinning of substrate-semiconductor interfaces considering DOS distributions either of square-root types or of Gaussian types. Our findings indicate that energy band at semiconductor interfaces with commonly observed unchanged $\varDelta_h$ (with $E_F$ pinning) is not only dominated by $\Phi_{sub}$ but also depending on film thickness with a relation of an exponential change. Furthermore, assumed constant charge transfer density in the depletion region of substrate-semiconductor interface should be carefully reconsidered since (i) a considerable amount of charge density may appear at the narrow region very close to the contacts, and (ii) the reduction of charge density to the neutral region with thickness could be much more significant than expected. Our calculations developed here shall provide a simple useful tool for quantitatively understanding the $E_F$ position in energy bandgap and formed hole/electron injection barriers at different types of semiconductor interfaces, thus paving the way towards the designing and engineering of novel semiconductor devices.

**Acknowledgement**

This work was financially supported by the Qing-Lan Project from Yangzhou University. The author declares no competing interests. All data is available in the main text or the supplementary materials.

**Reference**

[1] W. Schottky, Z. Phys. 113, 367 (1939).

[2] N. F. Mott, Proc. R. Soc. (London) A 171, 27 (1939).

[3] E. J. Mele and J. D. Joannopoulos, Phys. Rev. B 17, 1528 (1978)

[4] S. M. Sze and K. K. Ng, Physics of Semiconductor Devices, 3rd ed. (John Wiley and Sons, Hoboken, NJ, 2007).

[5] R. T. Tung, Appl. Phys. Rev. 1, 011304 (2014).

[6] W. Shockley, Bell Syst. Tech. J. 28, 435 (1949).

[7] S. Izawa, N. Shintaku, and M. Hiramoto, J. Phys. Chem. Lett., 9, 2914 (2018).

[8] H. Kanda, N. Shibayama, A. J. Huckaba, Y. Lee, S. Paek, N. Klipfel, C. Roldán-Carmona, V. I. E. Queloz, G. Grancini, Y. Zhang, M. Abuhelaiqa, K. T. Cho, M. Li, M. D. Mensi, S. Kinge and M. K. Nazeeruddin, Energy Environ. Sci., 13, 1222 (2020).

[9] S. Kurtin, T. C. McGill, and C. A. Mead, Phys. Rev. Lett., 22, 1433 (1969).

[10] A. Crispin, X. Crispin, M. Fahlman, M. Berggren, W.R. Salaneck, Appl. Phys. Lett., 89, 213503 (2006).

[11] Y. Liu, J. Guo, E. Zhu, L. Liao, S. Lee, M. Ding, I. Shakir, V. Gambin, Y. Huang &X. Duan, Nature, 557, 696 (2018).

[12] Z. Zhang and J. T. Yates, Jr., Chem. Rev., 112, 5520 (2012).

[13] M. Fahlman, S. Fabiano, V. Gueskine, D. Simon, M. Berggren and X. Crispin, Nature Reviews Materials, 4, 627 (2019).

[14] P. Schulz, D. Cahen and A. Kahn, Chem. Rev., 119, 3349 (2019).

[15] S. Krause, A. Scholl, and E. Umbach, Phys. Rev. B, 91, 195101 (2015).

[16] J. Yang, H. Sato, H. Orio, X. Liu, M. Fahlman, N. Ueno, H. Yoshida, T. Yamada, S. Kera, J. Phys. Chem. Lett., 12, 3773 (2021).

[17] I. Lange, J.C. Blakesley, J. Frisch, A. Vollmer, N. Koch, D. Neher, Phys. Rev. Lett. 106, 216402 (2011).

[18] H. Lee, B. G. Jeong, W. K. Bae, D. C. Lee & J. Lim, Nat. Commun. 12, 5669 (2021).

[19] A. Nigam, P. R. Nair, M. Premaratne, & V. R. Rao, IEEE Electron Device Letters, 35(5), 581 (2014).

[20] J. Bardeen, Phys. Rev. 71, 717 (1947).

[21] A. Cowley, & S. Sze, J. Appl. Phys. 36, 3212 (1965).

[22] H. Ishii, K. Sugiyama, E. Ito, K. Seki, Adv. Mater. 11, 605 (1999).

[23] S. Braun, W.R. Salaneck, M. Fahlman, Adv. Mater. 21, 1450 (2009).

[24] M. Oehzelt, N. Koch, G. Heimel, Nat. Commun. 5, 4174 (2014).

[25] P. Pingel and D. Neher, Phys. Rev. B, 87, 115209 (2013).

[26] M. L. Tietze, L. Burtone, M. Riede, B. Lussem, and K. Leo, Phys. Rev. B, 86, 035320 (2012).

[27] J. Yang, F. Bussolotti, S. Kera and N. Ueno, J. Phys. D: Appl. Phys. 50, 423002 (2017).

[28] P. Schulz, E. Edri, S. Kirmayer, G. Hodes, D. Cahen and A. Kahn, Energy Environ. Sci.,7, 1377 (2014).

[29] M. Yang and J. Yang, Appl. Phys. Lett. 117, 071602 (2020).

[30] V. Adinolfi, M. Yuan, R. Comin, E. S. Thibau, D. Shi, M. I. Saidaminov, P. Kanjanaboos, D. Kopilovic, S. Hoogland, Z.- H. Lu, O. M. Bakr, E. H. Sargent, Adv. Mater., 28, 3406 (2016).

[31] Z. Ni, C. Bao, Y. Liu, Q. Jiang, W. Wu, S. Chen, X. Dai, B. Chen, B. Hartweg, Z. Yu, Z. Holman, J. Huang, Science, 367, 1352 (2020).

**Table I.** Parameters used in the exemplary study with different types of DOS distributions.

| DOS type | $\varepsilon_r$ | $N_V$ (cm$^{-3}$) | $N_C$ (cm$^{-3}$) | $E_{H\text{-peak}}$ (eV) | $E_{L\text{-peak}}$ (eV) |
|---|---|---|---|---|---|
| Gaussian | 3.6 | 2.89×10$^{21}$ | 2.89×10$^{21}$ | 5.3 | 2.1 |
| **DOS type** | $\varepsilon_r$ | $N_V$ (cm$^{-3}$) | $N_C$ (cm$^{-3}$) | $E_V$ (eV) | $E_C$ (eV) |
| Square root | 3.6 | 1.07×10$^{19}$ | 1.07×10$^{19}$ | 4.9 | 2.5 |
| Silicon | 11.7 | 1.04×10$^{19}$ | 2.8×10$^{19}$ | 5.13 | 4.01 |
| MAPbI$_3$ | 17 | 2.8×10$^{18}$ | 1.9×10$^{18}$ | 5.4 | 3.7 |
| **Dopant** | $N_A/N_V$ (or $N_A/N_H$) | $E_A$ (eV) | $E_A$ (eV) | $\sigma_{H/L}$ (eV) | |
| Acceptor | 10$^{-8}$~10$^{-2}$ | 0.044 | 0.4 | 0.2 | |

**Figure 1.** Calculated results of $E_F$ change with respect to the valence band ($E_{V/H}$) at semiconductor interfaces considering different distances from the interface and various doping ratios, where exemplary substrate work function is 3 eV. (a) The $E_F$ changes as a function of doping ratio, where two types of DOS distributions (a square-root and Gaussian) are considered. The definition of hole injection barrier ($\Delta_h = E_F - E_{V/H}$) is given on the inset of (a). $\Delta_h$ changes as a function of film thickness under intrinsic and extrinsic semiconductors (four different doping ratios), where a square-root-type DOS in (b) and a Gaussian-type DOS in (c) are considered. The doping ratios in (b) and (c) are also marked with solid and open dots in (a).

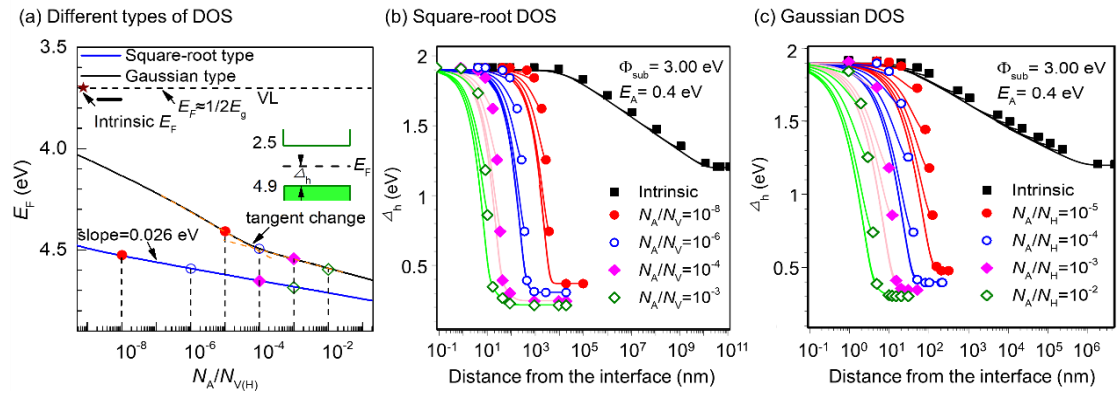

**Figure 2.** Calculated $\varDelta_h$ change under various substrate work functions, where a square-root type DOS in (a) and a Gaussian type DOS in (b) are considered. Both intrinsic and doped semiconductor interfaces are independently calculated under the same substrate work functions. Noticeable, $E_F$ shift demonstrates the relation to the distance from the interface is with an exponential type and any change of $\varDelta_h$ is limited in the shadow region marked in the figures.

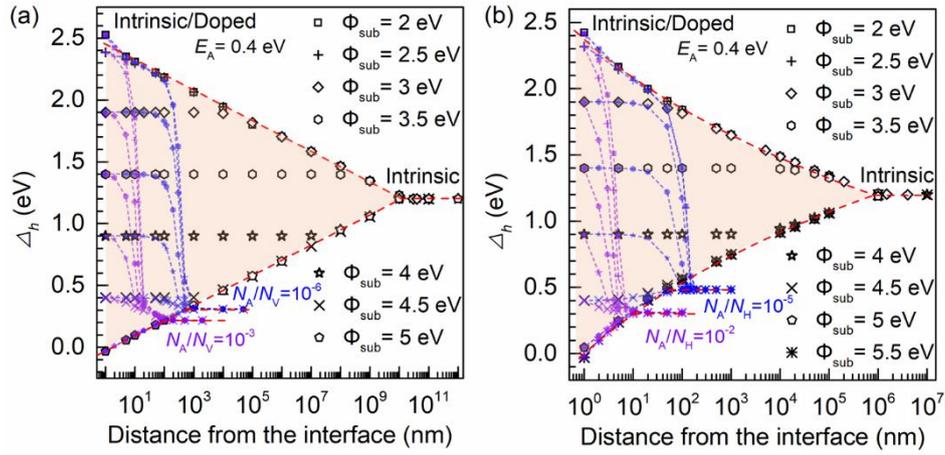

**Figure 3.** Calculated charge density (ρ) within various distances and substrate work functions in different doped semiconductors. (a) and (c) are exemplary results of square-root type DOSs; (b) and (d) are exemplary results of Gaussian type DOSs.

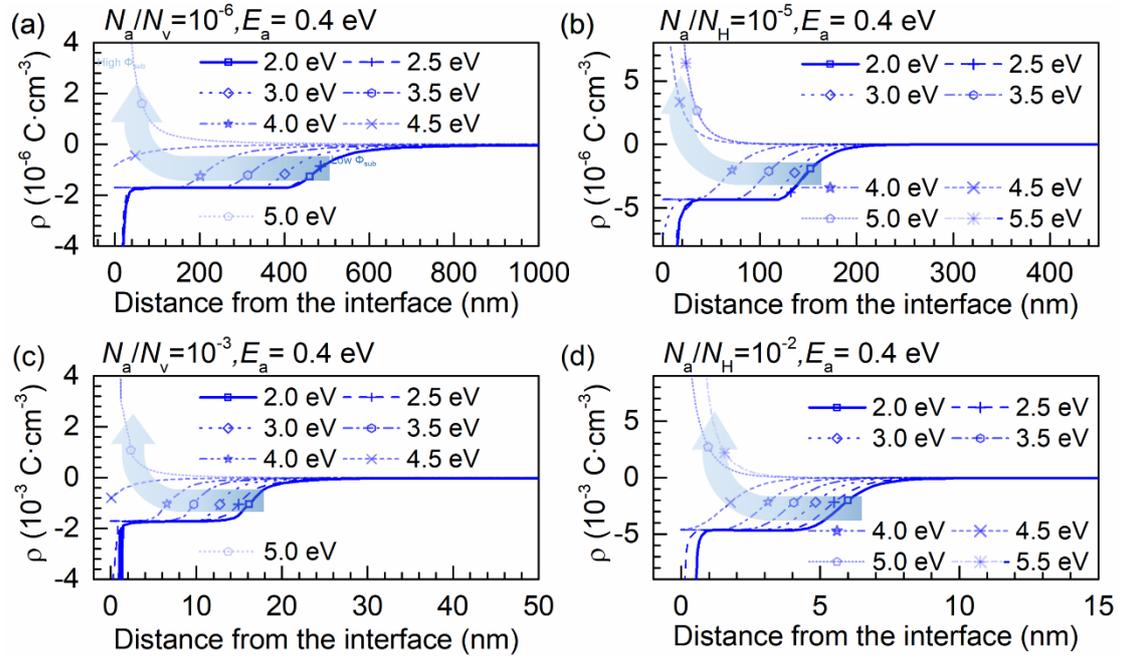

**Figure 4.** Two exemplary results giving typical semiconductor interfaces to specify our calculations, in which (a) is perovskite material (CH$_3$NH$_3$PbI$_3$) and (b) is a classical inorganic silicon.

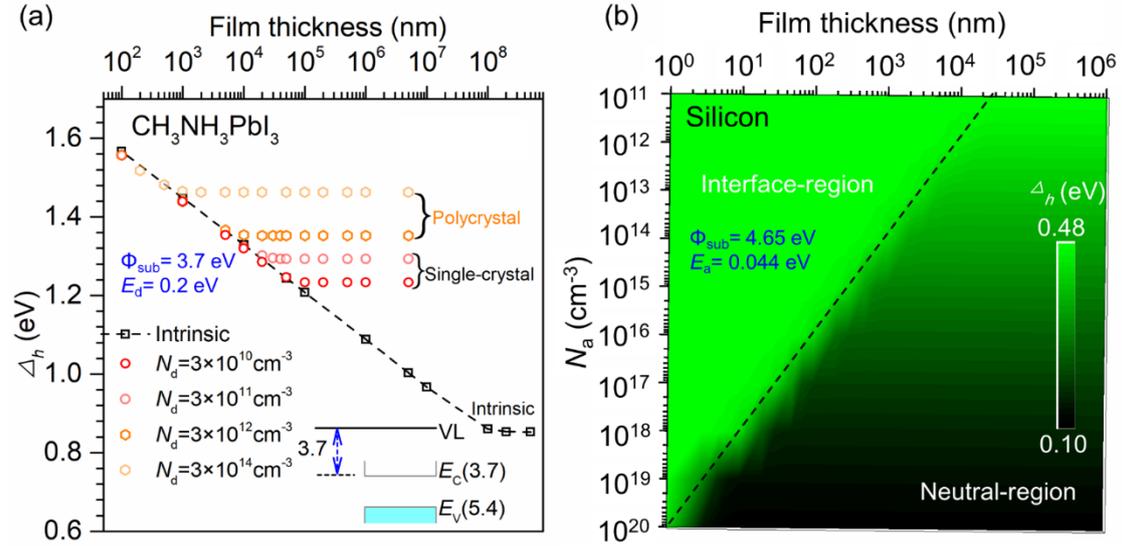

**Supplementary Materials**

**Disappearing of the Fermi level pinning at semiconductor interfaces**

Jin-Peng Yang[1*], Nobuo Ueno[2*]

*1, College of Physical Science and Technology, Yangzhou University, 225009, China;*

*2, Graduate School of Advanced Integration Science, Chiba University, Chiba, 2638522, Japan*

**1.** Detailed DOS distribution functions with $D_C(E)$, $D_V(E)$, $D_H(E)$, and $D_L(E)$.

$$D_C(E) = \frac{4\pi(2m_e^*)^{3/2}}{h^3}\sqrt{E - E_C} \quad E \geq E_c$$

$$D_V(E) = \frac{4\pi(2m_h^*)^{3/2}}{h^3}\sqrt{E_V - E} \quad E \leq E_V$$

$$D_H(E) = \frac{N_H}{\sigma_H \times \sqrt{2\pi}} e^{\frac{(E-E_H)^2}{2\sigma_H^2}} ;$$

$$D_L(E) = \frac{N_L}{\sigma_L \times \sqrt{2\pi}} e^{\frac{(E-E_L)^2}{2\sigma_L^2}}$$

Where $m_e^*$ is the electron effective mass, $m_h^*$ is the hole effective mass, $E_C$ is the conduction band edge, and $E_V$ is the valence band edge. Here, the energy gap ($E_g$) can be precisely given because DOS distributions strictly fall into zero. On the other hand, $\sigma_H$, $\sigma_L$, $E_H$ and $E_L$ are total density of HOMO states, total density of LUMO states, standard deviation of HOMO, standard deviation of LUMO, HOMO peak position and LUMO peak position, respectively. The $E_g$ in amorphous films is thus defined as the energy difference between the HOMO onset ($E_H - 2\sigma_H$) and LUMO onset ($E_L - 2\sigma_L$).

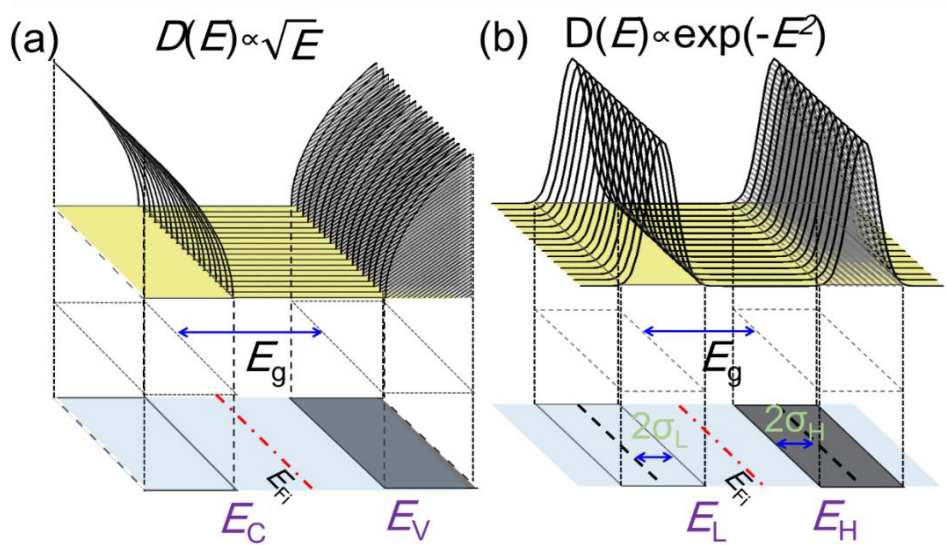

**Figure S1.** Illustrations of two different types of density of states (DOS) distributions at intrinsic semiconductors. (a) Square-root-type DOS with the conduction band ($E_C$) and the valence band ($E_V$). (b) Gaussian-type DOS with the lowest unoccupied molecular orbital (LUMO) and the highest occupied molecular orbital (HOMO). $E_L$ and $E_H$ are defined as onset energy position of Gaussian LUMO and HOMO peaks, which are shift with two times of standard deviation ($2\sigma_L$ or $2\sigma_H$). For intrinsic semiconductors, it is noticeable that $E_F$ position should be located in the middle of energy bandgap when effective masses of electron and hole are the same for square-root-type DOS and $\sigma_L$ is equal to $\sigma_H$ for Gaussian-type DOS.

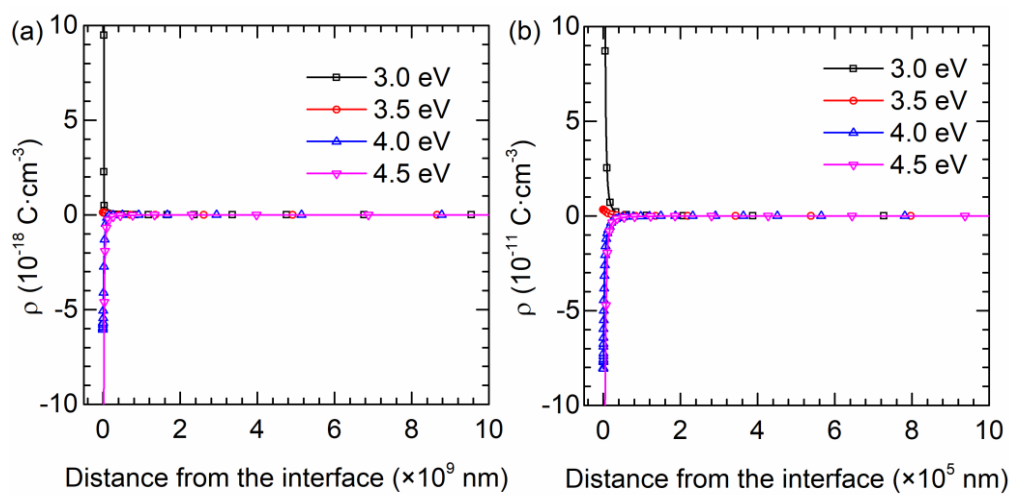

**Figure S2.** Calculated charge density distribution considering various distances and substrate work functions, where (a) is exemplary result of square-root type DOSs, and (b) is exemplary result of Gaussian type DOSs.

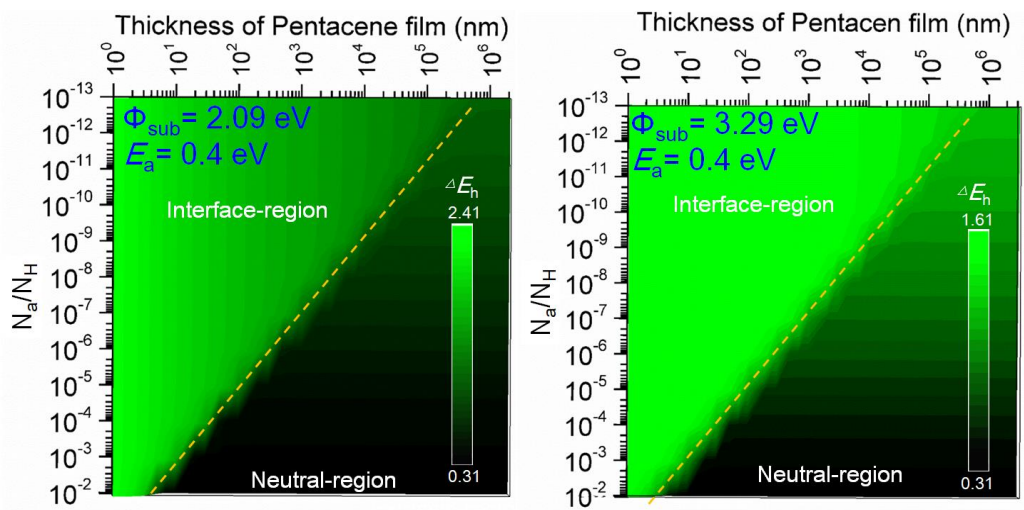

**Figure S3.** Two exemplary results giving organic (pentacene) semiconductor interfaces with considering their DOSs are Gaussian types. The parameters are obtained from Ref.1 [1].

**References**

[1] J.-P. Yang, L.-T. Shang, F. Bussolotti, L.-W. Cheng, W.-Q. Wang, X.-H. Zeng, S. Kera, Y.-Q. Li, J.-X. Tang, and N. Ueno, Org. Electron. 48, 172 (2017).